# Interdisciplinary Translation of Comparative Visualization

**James Birt**
Faculty of Society and Design
Bond University
Queensland, Australia
Email: jbirt@bond.edu.au

**Dirk S. Hovorka**
School of Business Information Systems
University of Sydney
New South Wales, Australia
Email: D.Hovorka@econ.usyd.edu.au

**Jonathan Nelson**
Faculty of Society and Design
Bond University
Queensland, Australia
Email: jnelson@bond.edu.au

## Abstract

Spatial visualisation skills and interpretations are critical in the design professions, but traditionally difficult to effectively teach. Visualization and multimedia presentation studies show positive improvements in learner outcomes for specific learning domains. But the development and translation of a comparative visualization pedagogy between disciplines is poorly understood. This research seeks to identify an approach to developing comparable multimodal and interactive visualizations and attendant student reflections for curriculum designers in courses that can utilize visualizations and manipulations. Results from previous use of comparative multimodal visualization pedagogy in a multimedia 3D modelling class are used as a guide to translation of pedagogy to architecture design. The focus is how to guide the use of comparative multimodal visualizations through media properties, lesson sequencing, and reflection to inform effective instruction and learning.

## 1   Introduction

As focus on technology enhanced teaching and learning (Keppell et al. 2011) and focus on awareness (Johnson et al. 2015a) and use (Johnson et al. 2015b) of emerging technology increases in higher education, there is an increasing potential in considering how technologies can enhance classroom pedagogy. Advances in technology have resulted in the development of new tools, techniques, and instrumentation that allow visualizations at different and multiple scales and the design and implementation of comparative pedagogy across multiple disciplines (Magana 2014). For example, the availability of low cost professional game engines and a range of affordable peripherals create new opportunities for the use of visualizations across classroom and discipline settings. In turn, visualizations (Freitas and Neumann 2009; Höffler 2010) can enhance the construction of knowledge, and provide a framework for collaboration which enhance learning outcomes (Garrison 2011). But if our goal is to move a learner from shallow to deep learning, we need to turn research attention to environments and pedagogies incorporating learning design (Ocepek et al. 2013), appropriate media properties (Dalgarno and Lee 2010) and lesson sequencing (Kozma 1994).

Multimedia visualizations and multiple learning modalities are learning-design support tools (Mayer 2005, 2008, 2014; Moreno and Mayer 2007) and it is understood that learners themselves have different styles, needs and capabilities (Freitas and Neumann 2009; Höffler 2010; Mayer 2005, 2008, 2014; Ocepek et al. 2013). However, most prior work has been formed around explanatory words and pictures (Ayres 2015) with less attention to complex learning environments such as interactive visualizations, games and simulations. In a pilot study by the authors (Birt and Hovorka 2014) using interactive multimodal visualization it was disclosed that no single mode of visualization was preferred by all students (Mayer 2014; Moreno and Mayer 2007). However, student's preferences of visual media changed over the course of study suggesting that students were not merely grasping the specific principles demonstrated by the visualization (e.g texture, lighting, perspective) but were also learning to pre-visualize and manipulate target objects "in their minds eye".

This research therefore seeks to identify an approach to developing multimodal and interactive visualizations and attendant student reflections (Wylie and Chi 2014) for curriculum designers in courses that can utilize comparative visualizations and manipulations. The goal is to guide the use of emerging comparative multimodal visualization as pedagogy through media properties, lesson sequencing, and reflection to inform effective instruction and learning thus shifting research away from whether technology, simulation or visualization affects learning. The question becomes how to translate a successful comparative pedagogy developed in multimedia modelling to other disciplines. Architecture design was selected as the subject of this research primarily because it is an accredited design discipline with learning outcomes aligned with the original pilot study. We focus on the process of translating the teaching principles and learning environment to construct a comparative pedagogy in architecture to explore how comparative learning can be achieved. We offer preliminary results from the ongoing pilot study in architecture in which we addressed two questions: 1): "How do learners perceive the comparative capabilities of visualization media to support learning?" and 2): "Do learners preferences for visualization technologies change with task or over time?"

## 2   Visualizations in Learning

Visualization of information has been successfully incorporated into learning in numerous academic disciplines (Freitas and Neumann 2009; Höffler 2010). Creating intuitive, navigable and manipulable forms of information, including images, manipulable and navigable Virtual Reality (VR) environments and physical representations provide a variety of 2d and 3d representations that can enhance learning and skills acquisition in part, because humans have evolved to comprehend spatial relationships among objects rapidly and effectively. Additionally, visualizations enable people to move between material reality, meaning objects they can see and touch, to abstract ideas and symbolic generalizations of objects and solutions that exist virtually or in their imagination.

To enhance students' conceptualisation, manipulation, application, retention of knowledge and their skills, visualizations must follow specific learning design (Mayer 2005, 2008, 2014; Moreno and Mayer 2007). In part, visualizations must prime the learner's perception, engage their motivations, draw on prior knowledge, avoid working memory overload through specific learning objectives, provide multiple presentation modalities, move learners from shallow to deeper learning and allow learners the opportunity to apply and build their own mental models (Hwang and Hu 2013; Mayer 2014). However, there are many challenges to developing visualizations including choosing between 2d and 3d interfaces, physical or virtual navigation, interaction methods, selecting an appropriate level of detail and availability of the visualization media (Hwang and Hu 2013). Paradoxically over use of visualizations over multiple representations may lead to reduction in deeper learning (Ainsworth 2014).

In the context of 3d modelling, spatial visualization and interpretation are important skills for novice designers to develop. These skills are involved in visualizing shapes, rotation of objects, and how pieces of a given design solution fit together. The ability to quickly, creatively and effectively interpret 3d spaces and forms from 2d drawings and the inverse, to reduce 3d ideas to 2d representations for communication purposes, is generally regarded as a hallmark of the profession (Wu and Chiang 2013). However, these skills generally require significant experiential development over the course of years and while experienced designers are adept at performing these translations there exists a communication barrier from instructor to learner due to this skills gap.

To assist with these challenges, technologies such game engines, 3d printing and VR are becoming available for use commercially and thus able to be incorporated into the classroom. The 2015 NMC Higher Education Horizon Report (Johnson, et al. 2015a) and Technology Outlook for Australian Tertiary Education Report (Johnson, et al. 2015b) specifically highlight these technologies as key educational technologies. VR technologies are mature, but the uptake in education has been hindered by cost, expertise and capability. This is now changing with the recent wave of low cost immersive 3d VR technology by vendors such as Oculus RiftTM (http://www.oculusvr.com/) and powerful interactive game engines such as Unity3dTM (http://unity3d.com/). However, there still remains an innate lack of physical haptic feedback that one gains through physical media manipulation (Fowler 2015).

3d printing offers a way to bridge the gap between the virtual and the real and creates and enables a haptic feedback loop for learners. 3d printing has seen an explosion in the past five years due to low cost fused deposition modeling (FDM) systems by makers such as MakerBotTM (http://www.makerbot.com/). 3d printing at its basic level uses an additive manufacturing process to build objects up in layers using plastic polymer. Although the process is slow, 3d printing creates direct links between a virtual 3d based model and the formation of an accurate, scaled, physical representation from that model (Loy 2014). This direct linking of object making to computer modelling changes the relationship of the learner to the making of the object and subsequent use.

## 3   Comparative Visualization as Pedagogy

The fundamental assumption(s) of comparative visualization are: that no technology offers a silver bullet for students to grasp specific concepts; multiple representations must take advantage of the differences between the representations (Ainsworth 2014); and students learn through a variety of approaches. This reflects the general proponents of blended learning approaches that long appreciated and advocated for multiple modes of presentation, delivery and content (Bernard et al. 2014). Disciplines whose subject matter are suitable for 3d visual presentations (e.g. medical anatomy, architecture, geography, chemistry and media/game design) benefit from the observation that multiple 3d modes of engagement can be reinforcing and synergistic.

This research builds on a prior pilot study (Birt and Hovorka 2014) which explored the effect of comparative mixed media visualization pedagogy on learning outcomes in multimedia 3d modelling design. A visual example of the comparative visualization pedagogy is provided in Figure 1.

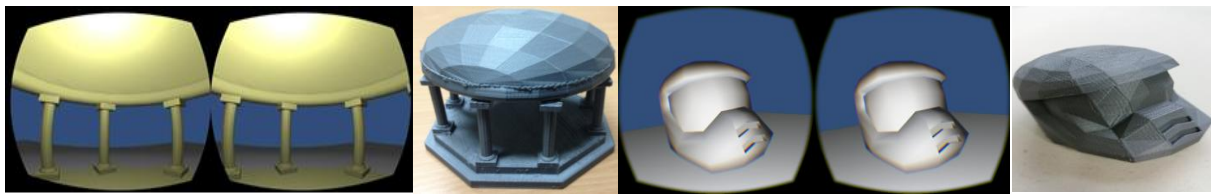

Figure 1: VR and 3d print visualization of 3d geometry learning objective

The learning artefacts shown in Figure 1 allow both high and low spatial learners to engage and conceptualize the object before constructing their own example. Additional examples are presented to the students, assisting construction of their mental model and application of theoretical geometry construction process. The learning objectives and resulting objects and their use in the classroom afforded learner centred active engagement through physical and virtual interaction with the visualization technologies. Research measures from each of the weekly learning objectives were achieved through coding and analysis of learner blogs conducted during the 12 week semester. Students were asked to engage in deeper learning and reflection by answering questions related to the weekly learning objective and technology visualizations. This included questions on: engagement; cognitive memory; visualization advantages/limitations; contrast between visualization media; how each technology would assist in demonstration of the learning objective to a team of designers; and communication of the learning objective between themselves and the instructor. The direct and reflective comparison between technologies revealed a strong interaction among them which enhanced the learning environment. Each visualization technology had positive, negative and mixed perceptions when it came to accessibility; usability; manipulability; navigability; visibility; communication; and creativity. The comparisons between delivery modes (visualization technologies) revealed that they provided much more than different versions of the same material. The engagement with each technology required reinterpretation of the principles upon which the lesson was focused. This provided students a way to "reframe" their own understanding and to "fill in the gaps" they observed using other media.

From the original pilot study there were several identified limitations and future research directions which have been addressed in this current research:

i.  sample population limited to a single class convenience sample of multimedia design students - to improve the significance of the sample the intervention and pedagogy was implemented a second time in the same subject code using a separate population sample in the following year (no participants were present from the first pilot study)

ii. low resolution virtual reality headsets reducing emersion - to help reduce issues related to the virtual reality media new higher resolution headsets were used increasing the resolution from the Oculus Rift DK1 at a resolution of 1280 x 800 per eye to 1920 x 1080 resolution per eye using the Oculus Rift DK2. Unfortunately due to the increase in resolution the computing power required by the higher resolution system caused significant issues with the processing frame rates with the visualizations dropping below the recommended 90 frames per second

iii. motion sickness from the virtual reality simulation frame rates – these issues unfortunately were amplified by using the higher resolution devices and not having access to powerful computing hardware capable of running the devices at the recommended 90 frames per second

iv. misinterpretation of the wording for the 2d orthographic view – students mistook this for the software tool -  regarding the misinterpretation several students still made incorrect linkage between the 2d orthographic views and the development tools used to construct the weekly tutorial exercises. This issue is likely driven by the students perceiving the 2d environment used in the tool as the theoretical orthographic views or, that, students often mistake the learning objectives and tasks as tool usage. To assist with this an introductory exercise was presented to the students highlighting all three media and specific advice was given regarding the 2d orthographic views

v. no inclusion of class wide communication and discussion about the technology – in relation to the communication an additional question was asked in each weekly blog specifically around communication and time was set aside for all students to discuss the exercises and include this as part of their learner blogs

vi. a perceived lack of object navigation and manipulation in the virtual reality environment – to improve navigation and manipulation additional tools were added to the virtual reality examples including the option to switch off the mesh renderer and display only the object wireframe highlighting the 3d object geometry and methods to scale and translate the objects in space helping with chunking and scaffolding of the learning objectives

Reflection on these observations and limitations of the presentation of modelling principles were refined for use in the multimedia class for 2015 and presented in Table 1.

| 2014 Modelling Class Principles | Modified 2015 Modelling class Principles | Applied Media Comparisons | | |
|---|---|---|---|---|
| | | 2d | 3d VR | 3D Phy |
| | Introduce the basic theoretical paradigms of 3d modelling | Y | Y | Y |
| Geometry | Apply geometry construction methods | | Y | Y |
| Curves | Demonstrate applied knowledge of curved surfaces from polygons primitives | Y | Y | |
| Material shaders | Different outcomes via material shader algorithms | Y | Y | |
| Texture Mapping | Apply texture mapping methods | | Y | Y |
| Lighting | Applied lighting theory | Y | | Y |
| Detail | Applied knowledge of Level of Detail | | Y | Y |
| | Demonstrate applied knowledge of presenting a complex scene and ability to reflect and synthesize the course | Y | Y | Y |

*Table 1.  Refinement of multimedia 3d modelling principles 2014 to 2015*

The major difference between the original and second study in multimedia was the inclusion of the learner perception primer for the students to help reduce the issues around 2d orthographic confusion and scaffold the users through the visualisation media. An additional final task was added that tested the students through a take home examination of the learning objectives used in the semester by requiring all students to create a 3d textured and lit object and tasking each student with a reflective essay examining the visualization methods and learning objective synthesis. A second major change was an increased emphasis on the theoretical underpinning of the principles. Thus the comparative visualizations were supported by a more in-depth articulation of the abstract principles during the tutorial sessions. The actual visualizations and the comparisons were maintained to allow comparison of the outcomes. The change was in the chunking of content and delivery of the tutorial with time given to the comparisons and communication between the students and the tutor.

All the multimedia learning objectives and applied media conditions were developed in accordance with the pair-wise comparisons used. The technology affordances and necessity for a dual coding method were addressed. By providing the opportunity for direct comparison of the media condition and asking immediate feedback in the form of learner blogs from each student, each class enabled students to recognise both what the specific learning objective was and to reflect on which media had a stronger effect on their understanding of the principles. This evaluation method satisfies the need to evaluate the design and is proposed to record learner observations, testing, and simulation experiences of the media conditions.

## 4   Comparative Pedagogy Translation

Traditionally spatial visualisation is difficult to teach but is a critical skill in the architectural profession. It is also generally accepted that the level of skill development in this area is correlated to the overall skill of a developing architectural designer, and is apparent in students' ability to communicate their ideas visually and verbally (Wu Chiang 2013). But while design and building technology becomes more advanced and pushes forward the complexity of possible design work, the development of spatial visualisation and interpretation skills in students runs the risk of being pushed aside in curriculum in favour of developing software and workflow skills due to the relative ease of teaching them. However, the problem persists that these abilities are articulated most effectively when in support of well-developed visualization skills.

The comparative pedagogy developed in the multimedia modelling classes focus on the principles to be learned, not the specific technology used with hopes to (i) gain insight into spatial visualization skills in students and (ii) contribute to knowledge regarding visualizations in learning and to explore translation of these concepts to new disciplines. Therefore, the challenge in translating from the multimedia pilot studies was to identify and abstract a set of core architectural principles which could be supported through the comparative visualization pedagogy.

For the translation to architecture, it was decided in early talks between the lead author of the multimedia study and the faculty from the architecture school that the comparative dual coding method approach would be integrated into the 'Architecture Design Communications' subject curriculum. Students in this course are exposed to traditional 2d hand drawing and digital 3d modelling and physical fabrication. Additionally, students are tasked with decomposing 3d virtual objects into 2d representations and reassembling these into 3d physical objects. Due to the nature of the assessment tasks and learning outcomes this closely aligned in terms of technology use and desired learning outcomes of the multimedia studies.

The mixed media comparison exercises and alignment with the second pilot study in multimedia addressed several of the future work outcomes of the original pilot study. These inclusions were included and developed on in close coordination with the architecture course designer. Curricular structure was tied directly to weekly lecture topics and tutorial sessions. As the classes are accredited specific objectives were required and this was considered throughout the design process.  Importantly, the translation process disclosed taken for granted principles for architectural learning and had not previously been emphasized. Table 2 highlights the weekly learning objectives and areas of disclosure for an architecture design course.

| Learning Objective | Primary Areas of Disclosure |
| --- | --- |
| Introduce the basic theoretical paradigms of 3d modelling | Increasing awareness of 3d visualization principles |
| 3d primitive construction and manipulation | Need to identify primitive from which complexity if constructed and understood |
| Knowledge of curves and NURBS surfaces | |
| Construction of complex surfaces | |
| 3d modelling as it relates to the human scale | Architecture is human – scale and visualizations require sense of place and space for human navigation |
| Management of complex scenes with a high number of models | |
| Navigation and management of Space | |
| Demonstrate applied knowledge of presenting a complex architectural scene | Whereas simplicity in multimedia modelling is a benefit, architecture is always situated and complex |

*Table 2.  Pedagogical disclosures of translation from multimedia to architecture design*

## 4.1　Disclosures from Multimedia to Architecture

The process of translation was an important mechanism for development of a comparative pedagogy. Specific attention was given to introduction of the techniques in the first week to reduce student confusion around the 2d orthographic framework and tool usage and scaffold the students across the various visualization tools as per the second multimedia study. This was also important as many assumptions were made regarding the knowledge-base of the multimedia students and their abilities to navigate virtual environments. Outcomes indicated that they were not familiar with gaming navigation and the use of keyboards/mice for virtual world navigation.

In discussions, the architecture faculty noted that, although the principles (e.g. lighting, texture, curves, etc) were quite similar the situated context of architecture required the visualization to place students differently. While multimedia modelling students benefit from manipulation of physical 3d objects and rotation of 3d models, architecture requires an understanding of navigation and space at a human scale. In addition, human structures are always situated, in that they are created and create a whole environment both internal and external. While this is not a problem per se for visualization it does change the actions and comparisons used in the pedagogy to enable students to grasp these principles. A summary of the differences disclosed in the translation process and an illustrative example of the first learning objective related to 3d modelling paradigms and theory is provided in Figure 2.

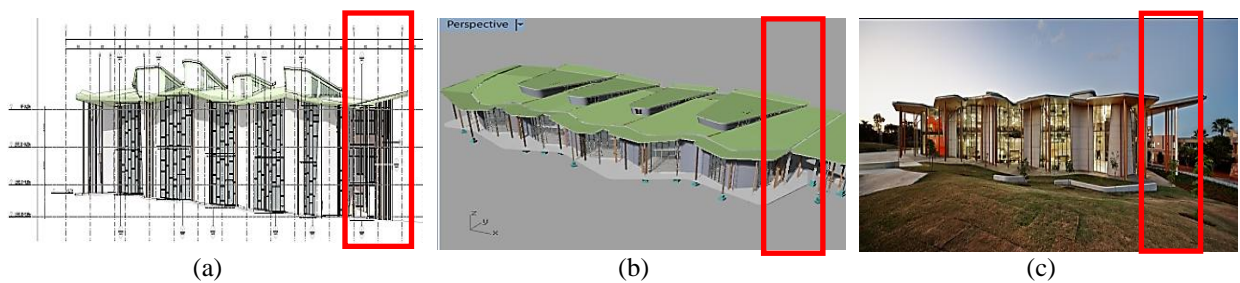

　　　　　　　(a)　　　　　　　　　　　　　　　(b)　　　　　　　　　　　　　　　(c)

*Figure 2: A geometrically complex structure shown in 2d, virtual 3d and physical 3d*

To address problem relevance, a technology visualization is constructed for each learning objective and media condition. In 3d architectural design as in 3d multimedia design, as spatial and geometric ideas become increasingly complex the industry standard 2d representations tend to convey less information about a design and how it is to be interpreted. Figure 2 illustrates this by showing: (a) 2d orthographic elevation drawing of a geometrically complex structure, (b) virtual 3d model perspective and, (c) the physical building. While the 2d representation is useful in showing a simplified general arrangement of the building elements, many 2d drawings are required to fully illustrate the complexities and form of the design. In particular the region marked in Figure 2 (a) is not readily discernible from this projected vantage point as can be seen in Figure 2 (c). The virtual 3d model, while it serves to inform a more complete view of the structure and geometric characteristics, contains little to no data about physical assembly, nor does it facilitate a piecemeal selection of information about the structure which is the goal of the 2d projections. The physical building shown in Figure 2 (c) provides haptic feedback and navigation but lacks internal transition within the geometry and ways to view the structure in its entirety. These differences in utility and comprehensibility therefore necessitate the need for neophyte designers to develop the skills to quickly and effortlessly switch back and forth between various media both cognitively and physically. Over the course of eight exercises students were asked to compare various forms of media including 2d, 3d print, built environments and 3d VR, culminating in comparison of all three. These exercises are intended to provide practical concept conveyance.

In addition, architectural learning requires students to address combinations of the principles in an environmentally situated scene. An illustrative example of the complex scene learning objective is provided in Figure 3.

The scene represents an interactive VR visualization and simulated lighting cycle of a physical built environment on the learner's campus highlighting complex shapes, surfaces, lighting and human scale. In addition to these tutorial exercises, students were also tasked with producing a detailed 3d model of a room they are familiar with, such as their bedroom, and all the objects in it in an effort to familiarise themselves with the basic software operations and workflows, over the course of the semester.

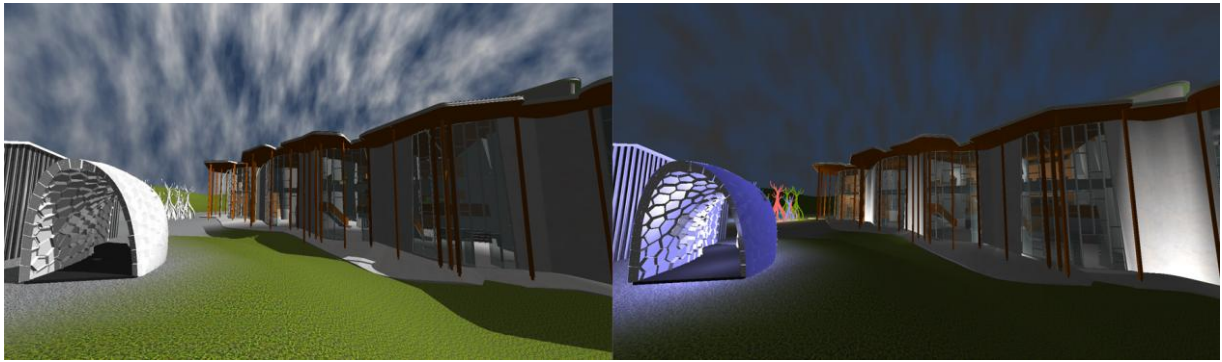

*Figure 3: A geometrically complex scene in VR of a physical built environment on the learner's campus at a 1:1 scale*

Each week students were given a milestone assessment blog which asked students to compare between different media representations tightly coordinated with lectures and the media experiments including (i) making rough 2d drawings of the room based on hand measurements of the actual room, comparing 3d physical and 2d projections , (ii) making rough 3d models from the 2d drawings and picture references, comparing 2d projections with 3d virtual, (iii) in an open critique and reflection session comparing side by side the 3d models and the reference pictures, comparing 3d physical to 3d virtual. Later in the semester, students were also allowed to visualise their design work in 3d augmented reality formats adding a new dimension to the ability to compare and contrast the media, with the goal being to allow them to learn to mediate between all three forms of representation and the physical full scale objects.

Further extrapolations of these objectives were also incorporated throughout the semester in an attempt to broaden the students' exposure to mixed media representations with different objects as deemed appropriate and to allow for repetition of skill development without repetition of subject matter. The semester ended with a final reflective session where students compared and contrasted the media between themselves, the applications of the media, and their current preferences for different tasks.

Translating from the multimedia pilot study, for each learning objective stimuli questions were used to capture the effects of the comparative pedagogy:

1. Which media representation(s) engaged you and what aspect(s) made it engaging?
2. Which media representation(s) did you find most memorable "sticky" in your understanding of the learning objective and why?
3. Discuss each media representation(s) advantages; limitations (constraints) and contrast the differences between each media representation.
4. For the purpose of demonstrating the learning objective to a design team, which media representation would you use and why?
5. Discuss with respect to the media representation(s) how the representation aided communication of the learning objectives between you and: your instructor; and your fellow classmates.

Preliminary analysis revealed themes resulting from the visualization comparisons. The themes with illustrative quotes are presented in Table 3. Further analysis must be conducted to enable correlation against student outcomes. This will allow for comparison with the first and second multimedia pilot studies to identify patterns of similarity and difference between the sets of students.

| 2d Image Projection | 3d Physical Built | 3d Virtual Reality |
|---|---|---|
| **Theme: Accessibility** | | |
| "orthographic drawings allow accessible exploration of ideas but lacks human scale" … "allows easy sharing of student experiences" | "this representation can be unavailable or difficult to navigate" | |

| **Theme: Usability** | | |
|---|---|---|
| "easy to use … drawing comes naturally" … "technical understanding of how pieces are erected on site" … "easy to fabricate" … "restricts scale of the project in relation to human scale" … "placing of each installation in relation to each" | "reality … understandable" … "most effective or easy to understand" | "initially unimpressed … hardware wasn't capable of smoothly rendering the types of buildings we were exploring in physical space … however improvements in hardware during the semester and viewing different sized models opened up possibilities" … "to assemble a physical model from a virtual version would be a time consuming process" |
| **Theme: Navigability** | | |
| "lacks navigation" … "lacks sense of depth and perception" … "angle of space is lost" | "walking around in reality has more effect" | "understand the circulation of the space for someone who does not know the physical room … great way to illustrate how the space can be occupied" … "explore places I have never been … see views of the building that usually cannot be seen" … "interact within the space and view the (structure) from angles that could not be projected" |
| **Theme: Senses** | | |
| | "you can touch the stairs, railings, walls, floors" … "gives the viewer a sense of material and touch" … "smell … sticks in my mind" | "lacks physical touch" … "no sense of smell" |
| **Theme: Visibility** | | |
| "it is one thing to look at something in 2d planar but the reality sometimes can be very different" … "lacks perception of the way the object interacts with the ground surface" | "physical space is ever present" … "quality" | "virtual reality allows one to gain a sense of human scale in the design … to immerse yourself in your project to understand the scale … allows fresh views like having only just experienced the physical space" … "connection to the land how the project sits on the ground" … "brought to attention many of the mistakes I made"… "highlighted which parts of the real thing I found important to convey" |
| **Theme: Communication** | | |
| "2d models on the computer screen cannot convey immersive and realistic impression of occupying space" | "most preferred method of demonstrating the learning objective to a designer" … "allows for discussion of finer detail and realism" | "effective means of communicating immersive and realistic impression of occupying a space … but not as good as physical" |
| **Theme: Creativity** | | |
| | | "made me more in tune to the physical structural elements of my physical space that due to familiarity I usually gloss over while occupying it … I found it fascinating it was barren without material texture but enthralling space … it showed the space as a whole rather than my occupation of a few physical elements" |

*Table 3: Summary of the themes drawn from the architecture student learner blogs*

## 5　Discussion and Future Research

The results of the comparative pedagogy developed for architecture (Table 3) broadly support the initial expectation that comparing different visualization modes is both engaging and beneficial to student learning. Students report becoming adept at 3d interpretation and mental conversion between 3d and 2d more easily than would be expected using traditional curricular means as highlighted in the following quote:

"We students are now able to rationalize and understand which concepts would be more efficient in certain situations … it definitely is making everyone an enquirer as everyone comes up with a different way of communicating the key idea through the different media styles and dimensions".

But comparing the architecture results with the results of the 3d modelling class revealed something surprising. When engaged with comparative visualizations, the architecture students stress quite different learning themes than observed in the multimedia modelling class. In particular, there is more emphasis on navigation in the physical and virtual space and of scale with fewer discussions on manipulation as highlighted by the multimedia students. We interpret this to suggest that the focus of multimedia is on representational objects and the way they are manipulated and interact. In contrast the focus in architecture is on placement of structures in an environment and the navigation within and around that structure. In addition, there is an overwhelming focus from the architecture students on the virtual reality technology compared to the multimedia focus on the physical representations and 2d. The architecture students concentrate more on space and scale where the multimedia students are more reflective on creativity, creation of real objects and from the manner in which dimensionality changes representation.

This study highlights that the process of translation of multi-mode visualizations into architecture disclosed important domain specific differences that inform pedagogy. For example, the original modelling class benefited from the material manipulability of physical 3D printed objects and the capability to rotate 3D visual models. Although some of the same principles (e.g curves, surfaces, texture, and lighting) are relevant in architecture, the scale at which they operate is quite different. Instead of manipulation and rotation of models, architectural pedagogy benefits from navigability, distance and perspective at a human scale, thus requiring visualizations allowing construction of complex surfaces from primitives, creation and navigation of complex scenes, multiple perspectives and the ability to experience visualizations at a human scale. Disclosure of these critical differences as highlighted in Table 2 led to a beneficial restructuring of the course material to highlight factors which have been previously obscured.

Future research seeks to gain insight into the effectiveness of wholly 3d and VR representations of the built environment that can potentially help move the design industry from a traditional 2D environment toward working in higher dimensions. Additionally, future research should further develop the pedagogy to incorporate new skills and disciplines framing body knowledge to develop a guideline of comparative visualization use in the classroom.

Finally, it is valuable to be reminded that the goal is pedagogy and achieving learning, not merely using the technology itself. In this case the reminder comes from an architecture student himself:

"When reflecting on each of the experiences and learning objectives as a whole and taking into account each of the comparisons for varying media, it is evident to see a recurring pattern. This is apparent when reading through each of (my) learning blog posts, that the same theme appears to be a common thread through each of them. Whilst we were comparing and contrasting the different media, to analyse and understand the strengths and limitations of each, it was evident that none was a 'Hero medium' and each had its place.  Whilst all were effective in different aspects, each of the mediums is effective in areas other couldn't hope to be. To be a truly proficient and effective architect, having an understanding on how and why to apply each of these media to the relevant learning task is the most indispensable tool."

## 6　References